\documentclass[AMA,LATO1COL]{WileyNJD-v2}
\usepackage{moreverb}
\usepackage{makecell}
\usepackage{amssymb}
\usepackage{multirow}
\usepackage{subfigure}
\usepackage{float}
\usepackage{xcolor}
\usepackage{array,ragged2e}
\usepackage{ulem}
\usepackage[T1]{fontenc}
\usepackage[utf8]{inputenc}
\usepackage{longtable}
\usepackage{tabu}

\usepackage{framed}
\newcommand{\highlight}[1]{\begin{leftbar}\noindent \emph{#1}\end{leftbar}}

\newcommand{\resq}{
Which qualitative methods, practices, and quality assurance frameworks from the social and behavioral sciences are currently underutilized in and could benefit software engineering research?}

\newcommand\BibTeX{{\rmfamily B\kern-.05em \textsc{i\kern-.025em b}\kern-.08em
T\kern-.1667em\lower.7ex\hbox{E}\kern-.125emX}}

\articletype{Article Type}%

\received{<day> <Month>, <year>}
\revised{<day> <Month>, <year>}
\accepted{<day> <Month>, <year>}

\begin{document}

\title{Qualitative Software Engineering Research -- Reflections and Guidelines}

\author[1]{Per Lenberg*}

\author[1,2]{Robert Feldt}

\author[1]{Lucas Gren}

\author[3]{Lars G{\"o}ran Wallgren Tengberg}

\author[3]{Inga Tidefors}

\author[4]{Daniel Graziotin}

\authormark{Lenberg \textsc{et al}}

\address[1]{\orgdiv{Department of Computer Science}, \orgname{University of Gothenburg}, \orgaddress{\country{Sweden}}}

\address[2]{\orgdiv{Department of Software Engineering}, \orgname{Blekinge Institute of Technology}, \orgaddress{ \country{Sweden}}}

\address[3]{\orgdiv{Department of Psychology}, \orgname{University of Gothenburg}, \orgaddress{\country{Sweden}}}

\address[4]{\orgdiv{Institute of Software Engineering}, \orgname{University of Stuttgart}, \orgaddress{\country{Germany}}}

\corres{*Per Lenberg, Corresponding address. \email{perle@chalmers.com}}

\presentaddress{Present address}

\abstract[Abstract]{
Researchers are increasingly recognizing the importance of human aspects in software development. Since qualitative methods are used to explore human behavior in-depth, we believe that studies using such methods will become more common.
Existing qualitative software engineering guidelines do not cover the full breadth of qualitative methods and the knowledge on how to use them like in social sciences.

The purpose of this study was to extend the software engineering community's current body of knowledge regarding available qualitative methods and their quality assurance frameworks, and to provide recommendations and guidelines for their use. 
With the support of an epistemological argument and a survey of the literature, we suggest that future research would benefit from (1) utilizing a broader set of research methods, (2) more strongly emphasizing reflexivity, and (3) employing qualitative guidelines and quality criteria.

We present an overview of three qualitative methods commonly used in social sciences but rarely seen in software engineering research, namely interpretative phenomenological analysis, narrative analysis, and discourse analysis. Furthermore, we discuss the meaning of reflexivity in relation to the software engineering context and suggest means of fostering it. 

Our paper will help software engineering researchers better select and then guide the application of a broader set of qualitative research methods.
}

\jnlcitation{\cname{%
\author{Lenberg P.}, 
\author{Feldt R.}, 
\author{Gren L.}, 
\author{Wallgren Tengberg L. G.},
\author{Tidefors I.} and 
\author{Graziotin D.}} (\cyear{2022}), 
\ctitle{Qualitative Software Engineering Research -- Reflections and Guidelines}, \cjournal{Journal of Software: Evolution and Process}, \cvol{2017;00:1--6}.}

\maketitle
\section{Introduction}
\label{lbl_introduction}

The importance of considering the people involved in the software development process has repeatedly been recognized by both researchers and practitioners~\cite{weinberg1971psychology, perry1994people, highsmith2002agile, lenberg2015human}. Still, studies concerned with behavioral aspects of software engineering are not in the software engineering mainstream. In terms of the number of existing publications and level of knowledge, the research area can still be considered young~\cite{lenberg2015behavioral}.

In young research areas, studies tend to adopt an exploratory approach, making qualitative methods a valid, or even preferable, option when analyzing software engineers' behavior~\cite{stebbins2001exploratory, fossey2002understanding, brown2006doing}. Qualitative methods are commonly used when one attempts to understand deeply the ways in which people act, think or feel~\cite{seaman1999qualitative, langley1999strategies}. 

Even if software engineering research traditionally has had a tendency towards quantitative methods, there are qualitative guidelines, e.g., Seaman~\cite{seaman1999qualitative}, Dittrich et al.~\cite{dittrich2007special}, Runeson and H{\"o}st~\cite{runeson2009guidelines}, Stol et al.~\cite{stol2016grounded}, Sharp et al.~\cite{sharp2016role}, Stol and Fitzgerald~\cite{stol2018abc}, and Hoda~\cite{hoda2021socio}. Even taken together, these guidelines do not cover the full breadth of qualitative methods and the knowledge on how to use them, as commonly found in, for example, the social sciences.

Moreover, different methods reflect different assumptions made by researchers about the world. The conclusions from disparate epistemological viewpoints can yield very different results \cite{willig2013introducing}. One example from social psychology is the study of self-esteem \cite{2016toh}. Not only can different conclusions be drawn, but different assumptions might even affect which research questions are posed and which answers can be found. As researchers, we must, therefore, answer a set of questions before we choose a scientific method: (1) What kind of knowledge do I aim to create? (2) What are the assumptions that I make about the (material/social/psychological) world(s) which I study? (3) How do I conceptualize the role of the researcher in the research process? What is the relationship between myself and the knowledge I aim to generate?~\cite{willig2013introducing}. 

For simplicity, let us divide the epistemological positions into (post-)positivist and social constructivist. The positivist argues that there are objective truths in the world that can be studied without the researcher influencing the participant in any way. It stems from the natural sciences, and applied to psychology, the belief is that the mind could be studied in the same way. The constructivist, on the other hand, assumes that reality is socially constructed and that there is always a subjective stance between the researcher and the participant. Instead, this perspective captures the participant's life experience to learn from research \cite{ponterotto2005qualitative}.

The purpose of this study was to extend the community's current body of knowledge regarding available qualitative methods and their quality assurance frameworks, and to provide recommendations and guidelines for their use. These additional methods are directly applicable, especially given the fact that different qualitative methods already used in software engineering (e.g., Grounded Theory) are also based on the same epistemological assumptions as the other methods we propose. 

To meet the purpose, we reviewed a sample of contemporary research against accepted social science standards. The social sciences have a long history of studying human behavior. We argue that software engineering researchers would most certainly benefit from utilizing their gained methodology skills and knowledge on how to conduct effective qualitative studies.

Based on the result of the review, we compiled customized recommendations and guidelines for future studies.

Our overarching research question is, hence,
\highlight{
\textit{\resq}
}

We acknowledge that, when taken in small doses, guidelines and practices can help to avoid more obvious errors and also help to frame qualitative work systematically and structurally~\cite{legreco2009discourse, tracy2010qualitative}. The quality of qualitative research does not rely solely on detailed guidelines and practices to produce sound research. Furthermore, a too intense methodological focus may risk creating anxieties that hinder creativity and practice~\cite{seale1999quality}. We encourage a reflection on the epistemological assumptions before the researchers collect empirical data.%

While we recognize the usefulness and advantage of a mixed model design~\cite{johnson2004mixed, creswell2013research} (i.e.,\ where the researcher combines a quantitative and a qualitative approach) we, for clarity, focus on qualitative methods only.

This paper is organized as follows. In the background section (\ref{lbl_background}), we give a brief introduction to qualitative research and  summarize the quality standards of qualitative research. Then (section \ref{lbl_literature_review}), we extract a sample of recent qualitative software engineering research and assess the quality and present our customized recommendations and guidelines.

In section \ref{lbl_reflexicity_and_methods}, we provide an in-depth presentation of the concept of reflexivity and present three qualitative research methods seldom used within software engineering research: interpretative phenomenological analysis, narrative analysis, and discourse analysis. Finally, we conclude the paper (section \ref{lbl_conclusion}). 

\section{Background}
\label{lbl_background}
In this section, we provide background information regarding subjects related to and relevant to this theoretical work. We first outline the conceptual framework in which we work, contribute, and present an overview of research in behavioral software engineering and qualitative research. We outline quality standards for qualitative research in general, and summarize quality standards for qualitative research in software engineering.

\subsection{Conceptual framework\label{ssec:framework}}

Stol and Fitzgerald~\citep{stol2018abc} report on inconsistencies in software engineering research regarding terminology in research methods and techniques. We agree with their stance; hence, we clarify some terms used in our paper. We make use of specific terminology that is derived from Creswell \& Creswell seminal book on research designs~\cite{creswell2013research}. When planning and performing a study, an interplay exists between research approaches, research designs, and research methods.

We consider a \textit{research approach} (alternative name: research methodology) to be the overarching philosophical perspective that underpins a researcher's beliefs and assumptions about the nature of reality, knowledge, and the way to gain knowledge. A research approach encompasses ontological, epistemological, and methodological commitments\footnote{We refer more commonly to \textit{worldviews} in the field of software engineering research, see, e.g., Petersen \& Gencel's paper on worldviews and research methods~\cite{petersenWorldviewsResearchMethods2013}}. Research approaches can be broadly classified into three categories: qualitative, quantitative and mixed methods. Research approaches allow us to prepare a blueprint for executing research. 

A choice for a \textit{research design} (alternative name: strategies of inquiries) is made within a research approach. A research design is the strategic framework that guides the empirical investigation of a research question. A design is a detailed plan outlining the specific steps to be taken to collect, analyze, and interpret data, guided by research questions, thereby ensuring that the study is methodologically rigorous and logically coherent.

A \textit{research method} refers to the specific techniques and tools employed to collect, analyze, and interpret data. A method is chosen based on its appropriateness for the research question and design, as well as its alignment with the researcher's ontological and epistemological assumptions. Research methods are about using questions to specify concrete steps for data collection, data analysis, interpretation, and validation.

In Creswell \& Creswell's framework, each component shapes the others. The research approach, being the most foundational element, influences the choice of research design and methods, ensuring that they are coherent and congruent with the researcher's philosophical stance. There is, however, not always a clear-cut between these categories. For example, Grounded Theory, in any of its variants, can be considered both a research design and a research method, typically within the qualitative research approach. The research design aspect of Grounded Theory provides a systematic framework for data collection and analysis, guiding researchers to inductively generate theory from empirical data. As a research method, Grounded Theory encompasses a set of concrete techniques and procedures for data collection, coding, categorization, and constant comparison, which are employed to analyze the data and generate emergent themes and concepts. Both components, but especially the research design one, vary according to the chosen Grounded Theory family~\cite{stol2016grounded}.

Interpretive Phenomenological Analysis (IPA), Narrative Analysis, and Discourse Analysis, which are presented in our paper, are mainly considered to be research methods within the qualitative research approach. These methods can be employed in, and adapted to, various research designs, depending on the research question, the objectives of the study, and researchers' stances on how they see the world.

Stol \& Fitzgerald~\cite{stol2018abc}, building on earlier work by Runkel and McGrath~\cite{runkel1972research,mcgrath1995Methodology}\footnote{Of note, Storey et al~\cite{storey2020who} also proposed to view (socio-technical) software engineering research from the perspective of Runkel and McGrath's work.}, presented a framework, named the \textit{ABC framework for software engineering research} to produce taxonomies of what they call research strategies, i.e., types of research methods. They distinguish between solution-seeking and knowledge-seeking studies. Solution-seeking studies are about building solutions for a software engineering challenge, typically in the form of software-related artifacts such as tools and models. Knowledge-seeking studies are about learning the world around us, in a software engineering context, including (software) systems, resulting artifacts, and users and developers. The ABC framework is for knowledge-seeking studies, which is also the scope of our contributions.

In their framework, Stol \& Fitzgerald place eight archetypal research strategies on a plane of two dimensions: the level of \textit{obtrusiveness} of the strategy in a given setting (including the level of settings manipulation caused by researchers) and  the extent to which the findings are \textit{generalizable} to other settings. Four quadrants categorize the research strategies:

\begin{description}
\item[Natural settings] existed before the researcher entered them. Archetypal strategies: field studies and field experiments.
\item[Contrived settings] set up specifically for the study. Archetypal strategies: experimental simulations and laboratory experiments.
\item[Neutral settings] set up to avoid affecting the research findings. Archetypal strategies: judgment studies and sample studies.
\item[Non-empirical settings] lacking empirical observations. Archetypal strategies: formal theory and computer simulations.
\end{description}

Where the neutral settings and the non-empirical settings meet, e.g., using sample studies or developing formal theory, there is potential for strongest generalization over actors (the A of the ABC). In the contrived settings, e.g., using a laboratory experiment, there is the highest potential for precision of measurement of actors' behavior (B in ABC). In natural settings, e.g., field studies, there is potential for maximizing the realism of context (C in ABC).
Stol \& Fitzgerald report, building on McGrath~\cite{mcgrath1995Methodology}, that maximizing generalizability jointly over the ABC (Actors, Behavior, and Context) is impossible since they are maximized for different types of research strategies of the framework.

Natural and neutral settings are opposed to each other along the generalizability scale. Studies in the neutral settings are about gathering data from systematically sampled participants who rate or judge phenomena (e.g., the Delphi method, or a focus group), as judgment studies or attempting to achieve generalizability within a population (e.g., a survey implemented as a questionnaire), using sample studies.
Studies in the natural settings can be either about observing in real-world context aiming for absence of manipulation (e.g., a case study based only on interviews), as field studies, or manipulating properties in the real-world context to observe an effect (e.g., action research), as field experiments.

In summary, we use Creswell \& Creswell's framework~\cite{creswell2013research} to contextualize research activities as well as for terminology. We use Stol \& Fitzgerald's ABC framework of software engineering research settings and strategies to contextualize our contribution to the software engineering domain~\cite{stol2018abc}.

\subsection{Research in behavioral software engineering}

There is a momentum in behavioral software engineering research, and methodological papers are delineating the discipline to grow sound and robust. The importance of considering not only technical aspects but the whole socio-technical system in which any piece of software will be situated has also received increasing attention, as evidenced for example by Storey et al.~\cite{storey2020who}. We are focusing here on our joint efforts in that we aimed to develop and offer three methodological papers to delineate the discipline. The readers will appreciate extensive references to works of others in the papers mentioned in this section.

A systematic literature review by Lenberg et al.~\cite{lenberg2015behavioral} indicated that the human aspect of software engineering is an important and growing area of research. However, the review also showed that there are knowledge gaps and that earlier research has been focused on a few concepts, which have been applied to a limited number of software engineering areas. Furthermore, this earlier research has rarely been conducted in collaboration by researchers from both software engineering and social science. 

Graziotin et al., \cite{graziotin2015}, meanwhile, offered a broader view of human aspects in software engineering research with a psychological perspective. The work offered concrete steps in selecting theoretical frameworks and psychometrically validated measurement instruments from psychology. Graziotin et al. called the field ``psychoempirical software engineering'' but later agreed with Lenberg et al. to unify the vision under ``behavioral software engineering''.

Hence, the second joint effort was born on quantitative behavioral software engineering. Graziotin et al.~\cite{graziotin2022psychometrics} performed an extensive survey of psychometric theory, framed by established standards, to develop and package introductory guidelines for software engineering researchers. The guidelines enable the adoption of existing valid and reliable instruments as well as developing new ones ``by the book''. The topics that we have touched upon are operationalizing psychological constructs, item pooling, item review, pilot testing, item analysis, factor analysis, statistical property of items, reliability, validity, and fairness in testing and test bias.

Gren~\cite{gren2018} offered a psychological test theory lens for characterizing validity and reliability in behavioral software engineering research, and he joined the present endeavor, continuing our shared efforts to establish the discipline. 

The present contribution is the third joint effort, on qualitative behavioral software engineering.

\subsection{An overview of qualitative research}
\label{lbl_qualitate_overview}
According to Corbin and Strauss~\cite{corbin2015basics}, qualitative research includes any study that produces findings that are not derived through quantification (e.g., statistical procedures). Quantitative research seeks causal determination, prediction, and generalization of findings~\cite{hoepfl1997choosing}. Qualitative studies, in turn, seek understanding and illumination of a certain phenomenon in a context-specific setting with the hope of extrapolation to similar situations. Such research addresses questions concerned with developing knowledge from the experience dimensions of human lives and social worlds. It aims to understand and represent the behaviors of people as they encounter, engage, and live through specific situations.~\cite{elliott1999evolving, patton1990qualitative}

To the best of our knowledge, no commonly accepted definition of qualitative research exists. Still, according to Lee et al.~\cite{lee1999qualitative}, such studies generally appear to have four defining characteristics. First, the data is derived from the participants' perspective, meaning that the researchers have not imposed a particular interpretation. It is the research participants’ subjective meanings, actions and social contexts, as understood by them, that is illuminated~\cite{fossey2002understanding}. The researchers attempt to bracket existing theory and their values, which allow them to understand and represent the participants' experiences and actions more adequately than would be otherwise possible.~\cite{elliott1999evolving}. Second, qualitative studies are often conducted in natural settings, whereas laboratory studies are rare. Third, in contrast to traditional, more rule-driven and survey-oriented approaches, qualitative studies are flexible and should be ready to change to match the fluid and dynamic demands of the immediate research situation~\cite{mack2005qualitative}. Fourth and final, no common standards for data collection and analysis exist, which may stand in contrast to prevailing beliefs about control, reliability, and validity~\cite{lee1999qualitative}. As previously stated, the qualitative research methods assume that there is at least a component of subjectivity in the research process, which is impossible to avoid \cite{willig2013introducing}. 

Qualitative research methods are useful in a variety of situations. Human behaviors, as individuals or in groups, are complex phenomena that often cannot be sufficiently described and explained through statistics and other quantitative methods. It thus calls for an alternative approach~\cite{seaman1999qualitative, langley1999strategies}. Qualitative methods are beneficial when addressing questions related to complex and versatile concepts such as behaviors, emotions, beliefs, and values. Additionally, such approaches can be useful when identifying latent and hidden factors whose role in the phenomenon under investigation may not be apparent (e.g., social norms, gender roles, or religion)~\cite{mack2005qualitative}. 

Qualitative approaches are also favorable when developing knowledge in poorly understood research areas~\cite{fossey2002understanding} and are therefore often used in exploratory studies~\cite{stebbins2001exploratory, brown2006doing}. A key is to pose open-ended interview questions, which provide the participants the opportunity to respond in their own words. Unlike closed questions, answers to such questions are not bound by the researcher's knowledge, but can stimulate responses that are meaningful and important to the participant.~\cite{mack2005qualitative} When using traditional quantitative data collection techniques (e.g., questionnaires), the researchers have no access to the reasoning behind the respondents' answers. Qualitative techniques, on the other hand, allow the researchers to better explore the underlying intrinsic processes.

Finally, qualitative methods can also be used to improve the validity of questionnaires and other survey instruments by extracting contextual data~\cite{fossey2001conceptual}. They can help to identify patterns and orders among variables and help to move inquiries toward more meaningful explanations~\cite{sofaer1999qualitative}.

In a special issue on qualitative research in software engineering, Dybå et al.~\cite{dybaa2011qualitative} claim that software engineering challenges can be difficult to study using quantitative methods. Qualitative research can be beneficial since software engineering research often includes small sample sizes. Moreover, the cost of controlled experiments with human subjects is high, and software engineering researchers need preliminary support before hypothesis testing can begin. Qualitative research in software engineering has, for example, been used to evaluate how agile software development affects teamwork~\cite{medeiros2020requirements} and to explore the consequences of using unclear design models when developing mobile software applications~\cite{farias2019designing}.

\subsection{Quality in qualitative research}
\label{lbl_quality_qualitative}
Criteria in qualitative research are, to say the least, challenging. If quality criteria are applied to qualitative research, which criteria  are appropriate and how should they be assessed has been up for debate for at least a quarter of a century~\cite{malterud2001qualitative, mays2007quality}. %

In general, scientists seem to hold three different opinions regarding the quality standard of qualitative work. Some argue that it makes little sense to attempt to establish a set of generic criteria since there is no unified qualitative research paradigm~\cite{rolfe2006validity}. Some claim that qualitative research can be assessed about the same broad criteria as quantitative research~\cite{malterud2001qualitative, mays2007quality}. Others suggest that, since qualitative research is based on different epistemological and ontological assumptions, the established criteria for scientific rigor in quantitative research cannot be applied to qualitative studies~\cite{lincoln1985naturalistic,chapple1998explicit}.

No matter what epistemological assumptions the researchers use, we argue that what constitutes sound research is of immense importance. Readers of scientific publications, researchers, and practitioners alike, need to know that the studies are trustworthy and provide solid findings, knowledge, and understanding of true events~\cite{fossey2002understanding, popay1998rationale}. The value of research is therefore, to a great extent, dependent on the researchers' ability to demonstrate the credibility of their findings~\cite{lecompte1982problems}.

According to Whittemore et al.~\cite{whittemore2001validity}, three of the most influential criteria have been outlined by Lincoln and Guba~\cite{lincoln1985naturalistic, guba1994competing}, Maxwell~\cite{maxwell1992understanding} and Sandelowski~\cite{sandelowski1986problem, sandelowski1993rigor} at the turn of the century. Lincoln and Guba~\cite{lincoln1985naturalistic, guba1994competing} propose five criteria for naturalistic inquirers: credibility, transferability, dependability, confirmability, and authenticity. Maxwell~\cite{maxwell1992understanding} further articulated the need for integrity and criticality, whereas Sandelowski~\cite{sandelowski1993rigor} advocated for creativity and artfulness.

Based on these criteria, Whittemore et al.~\cite{whittemore2001validity} suggest that one can divide qualitative research criteria into two categories: primary and secondary criteria. The primary criteria (credibility, authenticity, criticality, and integrity) are necessary to all qualitative inquiries, whereas the secondary criteria (explicitness, vividness, creativity, thoroughness, congruence, and sensitivity) provide further benchmarks of quality and are considered to be more flexible as applied to particular studies~\cite{whittemore2001validity}.

There are also been other researchers besides Whittemore that, in an attempted to identify the most important quality criteria, have aggregated and improved previous research. Drawing on previous research by Giacomini et al.~\cite{giacomini2000users} and Hammersley~\cite{hammersley1990reading}, Malterud~\cite{malterud2001qualitative} identifies relevance, validity, and reflexivity as three essential pillars of quality in qualitative studies. In addition, Elliot et al.~\cite{elliott1999evolving} claim that they built their criteria based on a list of more than forty different quality standards. The resulting list consisted of eleven principles: method appropriateness, openness, theoretical sensitivity (relating findings to existing knowledge), bracketing of expectations, replicability (describing methods), saturation generalizability (sampling adequacy for purpose), credibility checks, grounding (in examples), coherence, uncovering self-evidence to the reader and intelligibility (communicability).

Moreover, in two more recent studies, Tong et al.~\cite{tong2007consolidated} and Tracy~\cite{tracy2010qualitative} present two quality checklists for qualitative research. The latter presents and explores eight key markers of quality research---worthy topic, rich rigor, sincerity, credibility, resonance, significant contribution, ethics, and meaningful coherence. The former suggests a 32-item checklist grouped into three domains: research team and reflexivity, study design and data analysis, and reporting.

It is worth noticing that the quality standards mentioned above have primarily addressed, focused on, or been designed to consider medical or clinical applications qualitative work. Standards developed within a software engineering or even a work and organizational psychology context are considerably more scarce. There are, however, a few exceptions. In a study from 1999, Lee, Mitchell, and Sablynski~\cite{lee1999qualitative} reviewed qualitative studies of work and organizational psychology (WOP) for the past 20 years. Throughout the paper, the authors provide best practice recommendations. In addition, in a publication from 1997, Myers~\cite{myers1997qualitative} presents an overview of qualitative methods, addressed to the information system (IS) research community. 

\subsection{Qualitative research in software engineering}
\label{lbl_quality_qualitative_SE}
During the past twenty years, there have been a few method and process related papers addressing qualitative research in the software engineering domain. In an influential paper from 1999, Carolyn B. Seaman~\cite{seaman1999qualitative} introduces qualitative methods in software engineering. She argues that to further develop software engineering, new research methods are needed to explore non-technical aspects and that qualitative methods can be adapted and incorporated into the designs of empirical studies in software engineering. The paper presents an overview of methods for qualitative data collection and analysis. However, instead of presenting a broad spectrum of qualitative methods, Seaman has focused on detailing a selective few. For the data collection, she presents participant observation and interviewing, and for the analysis, she describes grounded theory. The author also briefly discusses threats to validity in qualitative studies and stresses the importance of considering triangulation, anomalies in data, negative case analysis, and replication.

Moreover, to our knowledge, there have been only two special journal issues dedicated to qualitative research in software engineering; one in the Information and Software Technology journal in 2007, and one in the Empirical Software Engineering journal in 2011. In an editorial to the former, Dittrich et al.~\cite{dittrich2007special} aimed to define qualitative research. Unlike Seaman, they clearly emphasize the diversity of qualitative research methods and also that qualitative studies are used under different epistemological orientations and with different theoretical underpinnings. 

Triggered by the inconsistency of reviews for the special issue, Dittrich and her co-authors took the first steps in developing a common way to evaluate the quality of qualitative research. Based on their experiences, they propose eight criteria for qualitative studies, emphasizing clarity of contribution of work.

Of the qualitative methods, grounded theory~\cite{glaser2009discovery} and thematic analysis~\cite{braun2006using} are the most popular with software engineering researchers~\cite{dittrich2007special, adolph2012reconciling, hoda2012developing, defranco2017content}. Since these are extensively used in software engineering research, we see no reason why other qualitative methods could not be applied directly to this field. They are based on the same epistemological assumptions and have been shown useful in other scientific fields~\cite{ponterotto2005qualitative}. 

The quality of software engineering studies using grounded theory was reviewed by Stol et al.~\cite{stol2016grounded}. Examining close to one hundred studies, the authors conclude that many papers do not generate a theory, do not clearly indicate which variant of grounded theory is used and do not provide sufficient methodological detail for rigorous evaluation. In addition, the authors present guidelines for how to conduct and report grounded theory studies. The guidelines are synthesized from existing methodological guidance and complemented with the authors own experiences. The guidelines, which are presented in the form of a checklist, are primarily directed to researchers with novice knowledge of grounded theory.

Finally, in a paper from 2016, Sharp et al.~\cite{sharp2016role} present the role of ethnographic work in software engineering. The authors argue that, despite its potential, ethnography has not been widely adopted in software engineering research. Their main aim was, therefore, to explain how software engineering researchers would benefit from adopting ethnography. They claim that the strength of ethnographic work is its ability to uncover the rationalities of the observed practices and that it, therefore, provides an important complement to other research methods that rely on a prior formulation of hypotheses. In addition, the paper introduces a guiding framework for ethnographic studies, supporting the design according to the research question being investigated, the context of the fieldwork and the characteristics of the main focus of the study. Again, if ethnographic studies have been shown useful in software engineering research, there is no reason why, for example, narrative analysis would not be applicable since they are both based on the same epistemological assumptions of investigating the lived experience of people \cite{harper1987visual}.

We believe that the method-related publications summarized in this section have contributed to increasing the knowledge of specific qualitative methods in software engineering. This is achieved by defining qualitative work, explaining how and in what way qualitative studies will contribute to the body of knowledge in software engineering, comparing qualitative and quantitative methods, and by presenting initial guidelines. Still, we notice that these publications have been focused on a few qualitative methods, and that these have been applied to a large extent using a positivistic epistemological underpinning. However, as we believe there is a component of subjectivity in the way a participant interprets her or his experiences and that a researcher plays a role in data collection in the creation of protocols, a constructivist standpoint would often make sense when these methods are applied. The descriptions and guidelines for constructivist-oriented methods do not exist, but they would be useful and help create a more nuanced picture of the current stage of research philosophy and method \cite{willig2007reflections}. 

\section{Literature review}
\label{lbl_literature_review}

As is stated in the introduction, it is part of our aims to assess the quality of software engineering qualitative studies, and also to identify common weaknesses. In this section, we extract a sample of recent studies and assess their quality against criteria defined in social science. 

\subsection{Method}
To identify a suitable sample, we performed a limited study of the literature.

As it is not the aim of our study to perform a systematic review of the literature, but to sense the field to provide an evidence-based justification for our guidelines, we looked into alternative proposals for literature review study designs. There are various design proposals for literature reviews that are placed somewhere in between the unsystematic--systematic pair and serve specific purposes~\cite{torraco_writing_2005,snyder_literature_2019}. While we do not identify our approach strongly with a particular one, we agree with Torraco~\cite{torraco_writing_2005}, who stresses how important it is to describe, follow, and report the process for selecting the articles. Hence, we adopted some steps proposed for systematic reviews, by Kitchenham~\cite{kitchenham2004procedures}, to enable both sound approach and reporting of it.

The processes included the following stages: selecting data sources, selecting search string, defining research selection criteria, defining the research selection process, and defining data extraction and synthesis. These stages, together with threats to validity, are presented in the following.

\subsubsection*{Data selection}

We limited the \textit{data sources} of the review to include peer-reviewed journal publications only. Peer-reviewed publications only as to meet a minimum threshold of quality that should be brought up by the review process. We decided to limit ourselves to journal papers to restrict the required effort.

Since qualitative research in software engineering can be considered an interdisciplinary research subject, we selected databases likely to cover both technical and social research, i.e., PsycInfo and Scopus.

Given our goal was not to perform a systematic literature review, we believe that these settings strike a fair balance between quality and representativeness of the gathered data and required effort. Yet, these settings bring important limitations that we discuss in section~\ref{lbl_limitations_review}.

The purpose of the \textit{search string} was to capture qualitative publications. Therefore, we combined ‘qualitative’ with synonyms to ‘system engineers’ (defined by Cruz et al.~\cite{cruz2011personality}). The final search string looked like this: \emph{("qualitative" OR "grounded theory" OR "thematic analysis" OR "discourse analysis" OR "narrative" OR "ethnography" OR "phenomenology" OR "case study" OR "content analysis") AND ("software engineering" OR "software development" OR "software engineer")}

To reduce the likelihood of bias, \textit{research selection criteria} were derived. The criteria were intended to identify those primary publications that provided insights relevant to the aim of the review.
\begin{enumerate}
\item[] Inclusion Criteria
\begin{enumerate}
\item[] \emph{Publication Year:} We limited the search to include papers published from 2015 to 2020.
\item[] \emph{Publication Type:} For quality reasons, we choose to only include peer-reviewed publications published in journals.
\item[] \emph{Content:} The publication shall use qualitative method(s) to study software engineering related activities or software engineers.
\end{enumerate}

\item[] Exclusion Criteria
\begin{enumerate}
\item[] \emph{Language:} We limited this study to only include papers written in English. Hence, we excluded all non-English publications. This, however, only applied to one publication.
\item[] \emph{Publication Type:} We excluded papers where we could not locate the full publication.
\item[] \emph{Content:} We acknowledge the usefulness and advantage of mixed model design~\cite{johnson2004mixed, creswell2013research}; however, for sake of clarity, we excluded mixed model studies. 
\end{enumerate}
\end{enumerate}

As for the \textit{research selection process}, the search identified 456 publications. We applied the selection criteria to the titles and abstracts and thus excluded papers that did not relate to software engineering related activities or software engineers. This reduced the number of potential publications down to 113. Then, we read the method section of the articles, which reduced the number of primary publications down to 61. 

\subsubsection*{Data extraction and analysis}
The aim of the literature review was to provide an overview of the current qualitative research in software engineering. To meet the aim, we extracted four properties from the included primary studies: (a) research method, (b) data collection method, (c) quality criteria indicators, and (d) quality guidelines. Information about these properties is presented in Table~\ref{table:properties}.

Regarding the quality criteria indicators, i.e., property (c), the choice of what quality criteria to use was not uncontested. We recognize that using the same set of criteria for all qualitative methods was not optimal, and that, to compile a detailed and nuanced assessment, we would have to apply different sets for different methods based on their underpinning epistemological orientation. However, we aimed not to compile a detailed assessment of each individual study. Instead, we strove to create general insights of the quality of the collective studies and we, therefore, argue that a common set of quality criteria for all types of research methods is sufficient. %

Moreover, as is stated in the introduction, we wanted to use quality criteria previously used in social science. Based on a general assumption that such criteria improve as the body of knowledge of qualitative research evolves and grows, we excluded criteria collections older than ten years. We also argue that a quality indicator, although weak, for such collections is its usage, which we estimated using citations.

Two criteria collections that met our requirements were the \emph{COREQ} checklist~\cite{tong2007consolidated} and the \emph{“Big-Tent”} criteria~\cite{tracy2010qualitative}. Our choice fell on the former for two reasons. First, we found the criteria in the COREQ to be easier to objectively assess compared to those in the \emph{“Big-Tent”}. As an example, in the \emph{eight “Big-Tent”}, one criterion is linked to how interesting the topic is, which clearly introduces a high degree of subjectivity. Second, we argue that the process used to compile the COREQ checklist is more structured and well-documented compared to the \emph{“Big-Tent”}. It should be noted, however, that some criteria in COREQ only appear to be applicable to interviews and focus group studies. Since our literature review indicates that these are the most commonly used data collection methods, that should not be a significant limitation.

The COREQ checklist, reproduced in Appendix C, consists of 32 criteria. Each criterion holds descriptive information in the form of guiding question(s). The criteria are grouped into the following eight themes: personal characteristics, relationship with participants, theoretical framework, participant selection, setting, data collection, data analysis, and reporting.

As an overarching goal, we strove to make the quality evaluation process as simple and straightforward as possible, thereby making it less affected by the researcher's prior knowledge, beliefs, and personal opinions. The result of the quality analysis of each criterion in the criteria collection was therefore binary, i.e., either the publication met the criterion or it did not. In general, a criterion was considered fulfilled if the publication provided an answer to the guiding question(s) associated with the criterion. We only assessed whether an answer was provided, not the quality of the answer. We clarify here that failing to provide an answer for a COREQ criterion does not necessarily entail that a publication lacked in quality. Authors were likely not aware of the existence of those criteria in the first place, which likely caused the missed explanation. All we can claim is that the publication did, or did not, provide an answer for each criterion.

As a consequence of this simplicity goal, we decided not to assess the criteria in theme \emph{reporting}, i.e., criteria number 29 to 32. We deemed that the guiding questions for these criteria required an in-depth analysis of consistency between aspects of the publication's content. For example, item 30 is ``Data and findings consistent---Was there consistency between the data presented and the findings?''. Such a criterion is hard to answer with a page number on which the answer appears at, which is what the COREQ checklist requires at bare minimum.

\begingroup
\footnotesize
\begin{longtabu}{ | p{0.2\textwidth}  | p{0.7\textwidth} | } 
\hline
\textbf{Property} & \textbf{Description and examples} \\ \hline
(a) Research method &
The research method comprises all processes that are used by the researchers during the study. Examples of research methods are grounded theory, thematic analysis, ethnography, phenomenology, discourse analysis, and narrative analysis. \\ \hline
(b) Data collection method &
Data collection is the process of systematic gathering or measuring information that enables the researchers to answer stated research questions. Qualitative data are varied in nature and can include any non-numerical information. Some major collection methods include interviews, focus groups, observation, and written texts~\cite{patton2005qualitative}. \\ \hline
(c) Quality indicators & 
Almost thirty quality criteria based on the COREQ checklist defined by Tong et al.~\cite{tong2007consolidated} were used. See table~\ref{table:coreq} for more details. We calculated the quality scores as percentages of all criteria across all included papers.\\ \hline
(d) Quality guidelines & 
We extracted information on whether authors have used any qualitative guidelines or checklists, e.g., 
\emph{COREQ} checklist~\cite{tong2007consolidated} or the \emph{“Big-Tent”} criteria~\cite{tracy2010qualitative}, to ensure the quality of their study. \\ \hline
\caption{Extracted properties}
\label{table:properties}
\end{longtabu}
\endgroup 

The analysis of the properties was straightforward and consisted only of a quantification and summarization of the extracted data. We determined the methods that were used and measured their frequency. The analysis of the fourth property (i.e., the quality criteria) was slightly more complicated. Since each of the thirty criteria was dichotomous and non-nuanced, we could not draw any decisive conclusions from the result of a single criterion. Instead, our findings needed to be drawn based on a cluster of criteria, all supporting the same result.

\subsubsection*{Limitations}\label{lbl_limitations_review}

The literature review sample was moderate, and we have not captured, as it was not our aim, all qualitative software engineering studies for the selected period. For example, our search string does not include all qualitative keywords, we are analyzing journal articles only, and we limited our search to be executed in Scopus and PsycInfo only. Hence, we cannot guarantee that some publications have mistakenly been excluded or missed. 

We are missing out on conference papers, which represent more than half of the total publication output in computer science, according to the DBLP\footnote{See \url{https://dblp.org/statistics/recordsindblp.html}.}. Qualitative studies are certainly published in conference proceedings but, we argue, studies that go beyond the mere qualitative labelling or coding of data find a better suited venue in journals, which are mostly not limiting submissions in any way, contrary to conferences. 

We use a limited amount of academic search engines. Scopus is one of the most used search engines and the subject of comparison studies with competitors such as Google Scholar and the Web of Knowledge, e.g., \cite{harzingGoogleScholarScopus2016,mongeonJournalCoverageWeb2016,s.adriaanseWebScienceScopus2013,wildgaardComparison17Authorlevel2015}. What these studies have found, among others, is that Scopus does not reach Google Scholar in terms of coverage, but it follows it closely, and has a better focus on publication quality. Furthermore, we enhance our coverage of publication with the American Psychological Association's PsycInfo, which is considered the major authority in the behavioral and social sciences.

We are also confident, since we adopted a sound and transparent methodology for our review, that we have not introduced systematic errors that could affect our findings. Our analysis has been careful, and we have only drawn conclusions when the data has been strong and unequivocal.

Moreover, the selection and the data extraction process were conducted by a single researcher. This approach is not as robust as having several researchers conducting the complete extraction in parallel.

Our literature review was aimed to motivate our contributions and to showcase the application of COREQ. We believe that it achieves its aims but want to highlight that it does not allow us to generalize regarding the whole research community.

\subsection{Result}
We have selected, extracted, and analyzed a total of 61 sources from the initial sample of 456. An overview of the extracted data is presented in Table~\ref{table:result_RR}. Appendix A shows all included papers, and a more detailed analysis of the quality scores is found in Appendix B (Empirical overview of the result for the quality indicator properties defined by COREQ) and in Appendix D (Papers that fulfilled each of the 28 first COREQ criteria)\footnote{Table~\ref{table:coreq2papers} in Appendix D provides a count of the papers for each criterion but omits the papers themselves, for legibility reasons. The full table is available online: \url{https://doi.org/10.5281/zenodo.7908320}}. 

\begingroup
\footnotesize
\begin{longtabu}{ | p{0.2\textwidth} | p{0.6\textwidth} | } \hline
\textbf{(a) Research method} & \emph{A Grounded theory} 44\% \newline \emph{B Not described} 30\% \newline \emph{C Thematic analysis} 13\% \newline \emph{D Other} 7\% \newline \emph{E Content analysis} 3\% \newline \emph{F Ethnography} 2\% \newline \emph{G Interaction analysis} 2\%  \\
\hline\emph{A}
\textbf{(b) Data collection method} & \emph{A Interview} 77\% \newline \emph{B Focus group} 11\%  \newline \emph{C Written text} 10\% \newline \emph{D Observation} 2\%. \newline In 46\% of the included studies, the researchers used more than one collection method. \\
\hline
\textbf{(c) Quality indicators}  %
& \emph{A: Personal Characteristics} 29\%  \\ 
& \emph{B: Relationship with participants} 14\% \\ %
& \emph{C: Theoretical framework} 40\% \\
& \emph{D: Participant selection} 50\% \\ 
& \emph{E: Setting} 44\% \\
& \emph{F: Data collection} 36\% \\
& \emph{G: Data analysis} 30\% \\

\hline
\textbf{(d) Quality guidelines} & No reference to any guideline or checklist was found. \\ 
\hline
\caption{Empirical overview of the result for the extracted properties.  The amounts are expressed as percentages of the 61 analyzed papers. For Research method property, the difference between "Other" and "Not described" is that in studies deemed as "Other", the method has been described, but it cannot be classified as any of the commonly used research methods.}
\label{table:result_RR}
\end{longtabu}
\endgroup

\paragraph{(a) Research method}
As is shown in the first row, software engineering researchers seem to prefer grounded theory as a research method, which was used in nearly half of the studies.

\paragraph{(b) Data collection method }
The most favored data collection method was interviews, which was applied in almost four out of five studies. It is also worth noticing that nearly half of studies collected data using more than one technique.

\paragraph{(c) Quality indicators}
The collected numerical data show that in most of the publications (70\%) the authors stated what research method they had used (theme \emph{theoretical framework} (C)). That means, however, that almost a third of the included papers did not state or describe the research method.

As the table shows, the theme \emph{relationship with participants} (B) had the lowest quality score. Worth noticing, and somewhat alarming, is that in less than ten percent of the studies the researchers discussed their assumptions (criterion 8 in Table~\ref{table:coreq}). This indicates that software engineering researchers seldom reflect on their bias (utilize reflexivity), or, at least, that these contemplations are not presented in the publications.

The quality score for \emph{personal characteristics} (A) was 29\%. In the papers that included the personal characteristics, this information was reported in the author's biography section. Thus, it was added since it was required by the journals, not as an active choice made by the researchers to raise the findings' credibility. These often brief presentations of the authors seldom provided any information about their qualitative research experiences.

As for the design of the studies, the statistical data shows that software engineering researchers frequently detail the duration of the focus group session or interviews, they describe what recording equipment was used, and they frequently present an interview guide. Nonetheless, our result indicates that data collection seems to be a one-off event rather than a continuous dialog. For example, interviews rarely were repeated (criterion 18), transcripts seldom were sent back to the participants for review (criterion 23) and the participants were rarely provided feedback on the findings (criterion 28).

In addition, a key feature in the most commonly used research method is data saturation, which means that researchers reach a point in their analysis of data whereby sampling more data will not lead to more information related to their research questions. Interestingly enough, saturation was only discussed in a third of the publications (criterion 22).  

Finally, most papers reported the number of participants (criterion 12) and presented some characteristics of the sample participants (criterion 16). More than half of them also provided at least some clues to how the participants were selected (criterion 10), but few presented how many refused to participate (criterion 13).

\paragraph{(d) Quality guidelines}
No reference to any qualitative guideline or checklist was found in any of the included publications.

\subsection{Discussion}
\label{lbl_discussion}

Based on our literature review, as well as our own research experience, we have identified three areas of improvement for qualitative software engineering research. We argue that software engineering qualitative research would benefit from (1) utilizing a broader set of research methods, (2) more strongly emphasizing reflexivity, and (3) employing qualitative guidelines and quality criteria. These areas are detailed in the following sections.

\subsubsection{A broader set of qualitative research methods}
According to social science researchers~\cite{willig2013introducing, smith2007qualitative, camic2003qualitative}, the most commonly used qualitative methods (and designs) are grounded theory, thematic analysis, ethnography, phenomenology, narrative analysis, and discourse analysis. Our review implies that among these methods, grounded theory, thematic analysis, and ethnography are established in the software engineering research community~\footnote{Readers may be puzzled by the absence of case studies, overall, in this paper. While a clear-cut categorization is not established all around, we consider, following Creswell \& Crewsell's framework (see Section~\ref{ssec:framework}), case studies as research designs---that is how they are sometimes referred to as, \textit{case study designs}---whereas the present paper mainly deals with research methods. To exemplify, we view a case study as a research design that employs one or more research methods, such as grounded theory, and gathers information using one or more data collection methods, like semi-structured interviews. Thus, a case study design may make use of, e.g., IPA, as a research method.}. This indication is further strengthened by the fact that the use and applicability of grounded theory and ethnography in software engineering context have, favorably, been reviewed and scrutinized in papers by Stol et al.~\cite{stol2016grounded} and Sharp et al.~\cite{sharp2016role}, and the fact that they are all more or less based on similar epistemological assumptions of reality.

Our review did not capture any publications that used phenomenology, narrative analysis, or discourse analysis. Social science researchers have, nonetheless, repeatedly recognized that these methods add value and that they are viable options when investigating organizational life~\cite{chia2000discourse, boje2004language, czarniawska1997narrative, feldman2004making, weick1995sensemaking, gill2014possibilities}. For example, Tomkins and Eatough~\cite{tomkins2013feel} suggest that phenomenology can be used to explore employees' meaning and experience of organizational life. Weick~\cite{weick1995sensemaking} advocates that stories or narratives are used to make sense of the complexity of organizational life, and they can hold information and influence organizational decision-making~\cite{martin1983uniqueness}. Finally, Chia~\cite{chia2000discourse} states that understanding organizational discourses is paramount for a deeper appreciation of the underlying motivational forces.

We, therefore, argue that phenomenology, narrative analysis, and discourse analysis also could contribute to the understanding of software engineering organizations and that much of the potential scope and value of these methods remain unrealized. We do not, however, claim that the researchers of the studies included in our review have chosen improper methods. We do not contend that the three methods should supplant existing well-established qualitative approaches. Rather, with a more varied toolbox to choose from, we believe that it would be possible to formulate complementary research questions and, possibly, highlight different aspects of software engineering phenomena, and, thereby, also yield more comprehensive insights.

\subsubsection{Emphasize reflexivity}
Our findings indicate that software engineering researchers seldom reflect on their assumptions and biases. Previous social science research has identified reflexivity as a crucial strategy in the process of generating knowledge by mean of qualitative research in general~\cite{gough2016reflexivity, malterud2001qualitative}. To raise the quality, credibility, and trustworthiness of qualitative research, it is important that the researchers, throughout the study, reflect on their opinions and how these affect their decision and the findings, but also that they report this in their publications.

We argue that the quality of qualitative software engineering research would improve if researchers would emphasize reflexivity. In addition to being a crucial strategy in qualitative research in general, we assert that reflexivity is of special importance in software engineering qualitative research. Such research is often conducted by researchers with a background in software engineering~\cite{lenberg2015behavioral}, which indicates that the researchers might have preconceived opinions of the phenomena under investigation and that the risk of research bias consequently is relatively higher.

In section \ref{section_reflexivity} below, we provide information regarding reflexivity and means of fostering it.

\subsubsection{Utilize qualitative guidelines or quality criteria}
We acknowledge that using guidelines or checklists as a mechanism to ensure quality might mislead qualitative researchers~\cite{reynolds2011quality}. What constitutes quality does not rely solely on detailed guidelines, and too much focus on checklists and processes risks creating anxieties that hinder creativity and practice~\cite{seale1999quality}. Still, since the software engineering research community, in general, is unfamiliar with qualitative studies, the benefits outweigh the risks. Taken in small doses, guidelines can help to guard software engineering researchers against more obvious errors and also help to frame qualitative work as systematic and structured~\cite{legreco2009discourse, tracy2010qualitative}.

Therefore, we recommend software engineering researchers to use the COREQ checklist as general guidance for ensuring quality at paper writing time. Providing details that cover most, if not all, the COREQ criteria will enhance the transparency of software engineering studies and how their results can be interpreted. We believe that quality could be increased even when explaining why certain criteria could not be covered, or if some do not even apply at all.
We deem that the criteria in COREQ are aligned with the software engineering research community's knowledge level and that they are relatively method-independent. Some criteria are, however, only applicable to interviews and focus group studies. Since our literature review indicates that these are the most commonly used data collection methods, that should not be a significant limitation.

For method-specific guidelines compiled for software engineering research, we recommend Stol et al.~\cite{stol2016grounded} for grounded theory, Sharp et al.~\cite{sharp2016role} for ethnography, Runeson and H{\"o}st~\cite{runeson2009guidelines} for case studies,
and Defranco et al.~\cite{defranco2017content} for content analysis.

\highlight{The COREQ quality criteria belong to the qualitative research approach and can be potentially applied to all qualitative research designs and methods. The ABC framework is not suitable to classify the COREQ criteria, but we suggest researchers operating in the natural settings as well to the neutral settings to observe the criteria and investigate their applicability in specific studies.}

\section{Reflexivity and Method Overview}
\label{lbl_reflexicity_and_methods}
In this section, we first provide information regarding reflexivity and means of fostering it. We then introduce phenomenology, narrative analysis, and discourse analysis, and highlight their limitations and challenges.%
\subsection{Reflexivity}
\label{section_reflexivity}
Reflexivity is regarded as a defining feature of qualitative research~\cite{gough2016reflexivity, malterud2001qualitative}. Still, even if its importance is recognized, to the extent of our knowledge, no commonly accepted definition exists. It is, however, commonly accepted that reflexivity is based on a recognition that the researchers are part of the social world that they study~\cite{palaganas2017reflexivity}, and that a key aspect is to make the relationship between the researcher and the participants as explicit and transparent as possible~\cite{palaganas2017reflexivity}. 

In qualitative studies, the researcher is considered the primary instrument of data collection and analysis~\cite{gough2012subjectivity, malterud2001qualitative}. The researchers' body of knowledge can thus be utilized to gain new and more in-depth insights of the phenomena under study.

Reflexivity involves reflecting about how our thinking came to be and how pre-existing understanding is constantly revised in the light of new insights~\cite{watt2007becoming, morrow2005quality, malterud2001qualitative}. It entails awareness that the researchers' involvement affects the research process~\cite{watt2007becoming} and could be viewed as a state of being that permeates all research phases, including the formation of research questions, data collection and data analysis~\cite{guillemin2004ethics, bradbury2007enhancing, berger2015now}. 

In a study with effective reflexivity, the researchers can also treat themselves as objects of inquiry~\cite{smith2006encouraging}. A reflexive researcher is sensitive to the ways in which she or he and the research process have shaped the collected data. Personal and intellectual biases need to be made plain at the outset of any research reports, enhancing the credibility of the findings~\cite{mays2000assessing}.

According to Russel and Kelly~\cite{russell2002research}, the absence of reflexivity may lead to acceptance of what is apparent and thereby obscure unexpected possibilities. In addition, if reflexivity is thoroughly maintained, personal opinions can be valuable sources for relevant and specific research.

Previous research has highlighted three main advantages of reflexivity. First, it is used to raise the trustworthiness of the study by making it more open and transparent; this is achieved by identifying and reporting the researchers' values, beliefs, knowledge, and biases~\cite{buckner2005taking, macbeth2001reflexivity, berger2015now}.

Second, reflexivity enhances the quality of the research by letting researchers reflect on who they are and their relation to the phenomena, which may both assist the process of constructing new insight~\cite{berger2015now}. However, the investigator should ensure not to confuse knowledge intuitively present in advance with knowledge emerging from the material in the study. Such situations can possibly be avoided by declaring beliefs before the start of the study~\cite{malterud2001qualitative}.

Third and final advantage, reflexivity helps to keep the research process ethical by helping to address concerns regarding negative effects of power in the researcher-to-participant relationship~\cite{berger2015now, pillow2003confession}. It helps maintain the ethics of the relationship between researcher and participant by equalizing their status, and ensures that the interpretation of the findings is always done through the eyes and cultural standards of the researcher~\cite{frisina2006back, josselson2007ethical}.

Even though the benefits of being reflexive are significant, it also comes at a cost for the researchers. It forces them to be transparent and expose their flaws, inner thoughts, and reasoning, which, potentially, could cause embarrassment or even shame. To protect themselves from being unmasked in the public research arena, scientists are reluctant to include human emotions and in-depth self-reflection in their publications~\cite{smith2006encouraging}.

We argue that software engineering researchers' usage of reflexivity is, at least partially, affected by the culture and norms of their various peer-groups, e.g., their research team, their university, and the software engineering community at large. An environment, where the researchers feel that it is safe to open up without being exploited, creates conditions that foster reflexivity. In a culture where genuineness and authenticity are the norm, scientists will certainly be more willing to reflect and communicate their honest thoughts, reasoning, and true feelings. The opposite is, however, also true. An environment where thoughts and feelings are considered signs of weakness that could be held against you clearly does not facilitate a reflexive behavior among the researchers.

True reflexivity is a state of being and cannot be encouraged only through guidelines and quality checklists. Reflexivity should permeate all aspects of the qualitative research and needs therefore to be a natural part of the software engineering researchers' professional identity. We thus claim that raising the level of reflexivity in qualitative software engineering research is clearly a community-joint effort. This calls for changes in several areas, e.g., mandatory courses that are included in the Ph.D. education to how the reviews of qualitative research are conducted.

Furthermore, it could also be useful to consult a research team or peer for dialog and discussions. They can serve as a mirror, reflecting the researchers’ responses to the research process, or they may also act as devil’s advocates and propose alternative interpretations to those of the researcher. Other strategies for maintaining reflexivity include repeated interviews with the same participants, triangulation, and peer review~\cite{berger2015now, morrow2005quality, frisina2006back, russell2002research, bradbury2007enhancing, padgett2016qualitative}.

\subsubsection{Group reflexivity}
Reflexivity is presented in the literature primarily as an activity carried out by single individuals~\cite{barry1999using}. Yet, qualitative research is often a team endeavor. The question on whether reflection is possible to be conducted as a team effort, how to do it, and how to report it, is justified~\footnote{In our conducted studies with IPA in the software engineering domain, in form of a research paper or a student report, the question on how to perform reflexivity in a research team has come all the time. This section was originally not present in the paper when submitted. We are thankful to an anonymous reviewer for their request to explore group reflexivity.}. 

Surprisingly, the literature on the matter is limited. Perhaps, as argued by Smith~\cite{smith2004reflecting}, the reason is that it is not an easy process to put the self in research endeavors, let alone one of a group reflection. Smith evaluated reflexivity frameworks and discovered that reflexivity in group settings is mentioned as a possible practice, but the \textit{how} is missing. One option might be the one of action learning cycles, as suggested by Pedler and Abbott~\cite{pedler_facilitating_2013}. Learning cycles is a framework to organize learning from experience, which resembles the action research cycle. The framework provides a set of questions to be asked during the reflection step, e.g., 
``What assumptions did we make?''. The framework, however, does not provide details on concrete steps, including how to report them. 

One approach to organize team reflexivity would be to mimic what others have done. For example, Palaganas et al.~\cite{palaganas2017reflexivity} provided a concrete example of group reflection: four authors state their backgrounds separately, then provide shared principles that touch upon ontological and epistemological issues and issues at politics, ethics, and personal value levels. 

Probst and Berenson~\cite{probst_double_2014} followed a similar approach in their study (detailed below): they stated their backgrounds, values, and interests separately first. Then, they reported their common goals and practices for the study. Finally, they ended with a further reflection by the first author on realizations that happened during the interview.

Driven by this lack of established practices, other researchers embarked on the empirical path to allow group reflection practices to emerge from data.
Barry et al.~\cite{barry1999using} expressed reporting a real-world study (for reflexivity carried out as team activity) mixing individual and group reflexivity and a dialogue between the two. They describe the team structure and dynamics, including the history of collaboration, people joining, tensions, geographical distribution commitment in the analysis of data. 

Barry et al.~\cite{barry1999using} provide two tools to be used to guide group reflexivity. The first tool is a set of seven orienting questions (e.g., ``What is my stake in the research?'', ``What do I hope to get out of it?''; ``In what way might my experience color my participation in the project?'') to be distributed in a first full team meeting. The authors reflect that this operation should rather happen after some initial bonding social activities, e.g., shared meals. 
By sharing the answers with each other, they report, the team reached an increased understanding of each other's positions. In particular, on the differences. The authors welcomed different values, way of thinking, expectations to avoid a lack of self-questioning. The value lies in disclosing and discovering differences before conducting the study. Furthermore, highlighting differences enables discovering the overlaps in worldviews.  

The second tool provided by Barry et al.~\cite{barry1999using} is to be used before starting to report research. All authors wrote their definitions for the central conceptual issues of the project. This would ground participants in the theory, coming from the literature. Differences emerging from these definitions led to valuable discussions and, hence, more conceptual analysis of the data.  

Probst and Berenson~\cite{probst_double_2014} interviewed 34 social work researchers to investigate their reflexive activities during research. Participants described ``a wide array of activities'' and ranges in terms of formality and collaboration techniques. Their results provide an entry for reflexive practices with others. In group reflexive actions, participants shared observations, questions, dilemmas, and difficulties with team members over different communication channels (from formal group meetings to e-mails and online chats). Reflexive actions were reported to happen initially at the individual level in pre-writing, ongoing writing, and post-writing phases, most frequently on logs, diaries, and field notes. The study, observing the diversity in techniques, expectations, behaviors, and attitudes,  concludes that the assumption that reflexivity needs to be more precisely mapped and operationalized---in our case in a group effort---may in the end be as problematic, although in different ways, than the need it attempts to resolve. 

The authors continue addressing a fundamental concept that we sought explanations for, namely, that mixing epistemologies by systematizing and  objectifying a fundamentally subjective process will not enhance rigor in the struggles to understand human experience. In other words, and quoting the abstract, 
``ultimately the mechanism of reflexivity may not lie in the specific activity but in the attitude with which it is carried out''. We suggest behavioral software engineering research to conduct group reflexivity in the ways that make the researchers most comfortable with, and to report such activities for transparency, and to inspire future studies.

\highlight{Reflexivity is influenced mostly by research approaches and philosophical assumptions rather than specific designs and methods. We see its applicability, in terms of the ABC framework, for strategies in the natural settings for the most part. Reflexivity is a tool that provides further details to maximize the C of the ABC, that is, realism of context.}

\subsection{Method overview}
\label{section_method_overview}
In an attempt to bolster the interest and broaden the range of qualitative research methods used by the software engineering research community, we describe three qualitative research methods---interpretative phenomenological analysis, narrative analysis, and discourse analysis. Thereby, enabling researchers to make well-founded decisions regarding choice of methodology.
Table \ref{lbl_method_overview} provides a brief overview of these methods and also guidance to when they could be applicable. With the ambition to raise the interest and the curiosity of these qualitative methods. In the following sections, we present a somewhat more detailed description and discuss their limitations and challenges.

\begingroup
\footnotesize
\begin{longtabu}{ | p{0.20\textwidth} | p{0.20\textwidth} | p{0.20\textwidth} | p{0.15\textwidth} | }
\hline
& \multicolumn{1}{c|}{\textbf{Phenomenology}} & \multicolumn{1}{c|}{\textbf{Narrative Analysis}} & \multicolumn{1}{c|}{\textbf{Discourse Analysis}} \\ \hline
\textbf{When to use?} 
& When interested in how software engineers make sense (experience) of a specific phenomenon in a given situation.
& When interested in how software engineers create meaning in their lives as stories (narratives). Compared to IPA, use narrative inquiry when you are interested in how a chain of experiences is weaved into a narrative, not the experience by and of itself.
& When interested in exploring social, well-established meanings or ideas around a topic that shape how software engineers can talk about it. To uncover how language is used to accomplish personal, social, and political projects.
\\ \hline
\textbf{Research questions (examples)} 
& R1. How does a software engineer experience organizational loyalty? 
R2. How do people make the decision to become a software engineer?
& R1. How do individual software engineers come to know their experience of the changes in ways of working that followed the introduction of agile methods? 
R2. What is the senior software engineer's story of the experience of transferring to a team-based organization? 
& R1. What discourse exists in software engineering organizations, and how do they empower some groups or roles while dis-empowering others?
R2. How do software engineers construct team-identities within agile teams?
\\ \hline
\textbf{Research outcome (examples)} 
& R1. An in-depth description of the essential structure of organizational loyalty experience. 
R2. A portrayal of what factors and experiences that shape and influence the decision. 
& R1. Narrative that accounts for software engineers' experience of changes in their ways of working.
R2. Narrative of a senior software engineers' experience of transferring to a team-based working environment.
& R1. Description of how different discourses shape relationships and how social goods are negotiated and produced in a software engineering organization.
R2. A summary of discourses that affect and  strengthen or weaken team-identities in agile teams.
\\ \hline
\textbf{Reference studies} 
& \emph{The transition to motherhood in an organizational context: An interpretative phenomenological analysis} by Millward~\cite{millward2006transition}; 
\emph{Elite identity and status anxiety: An interpretative phenomenological analysis of management consultants} by Gill~\cite{gill2015elite}.
& \emph{I am not a tragedy. I am full of hope’: communication impairment narratives in newspapers} by Malley ~\cite{malley2014not}; 
\emph{Complexities of identity formation: A narrative inquiry of an EFL teacher} by Tsui~\cite{tsui2007complexities}.
& \emph{Cognitive organization and identity maintenance in multicultural teams: A discourse analysis of decision-making meetings} by Aritz and Walker~\cite{aritz2010cognitive}; 
\emph{Articulating circumstance, identity, and practice: toward a discursive framework of organizational changing} by Jian~\cite{jian2011articulating}.
\\ \hline
\caption{The table presents an overview of qualitative methods, and it provides examples of applications of these methods in software engineering.} %
\label{lbl_method_overview}
\end{longtabu}
\endgroup

\subsubsection{Interpretative Phenomenological Analysis}
The purpose of interpretative phenomenological analysis (IPA) is to explore in-depth the processes through which people make sense of their experiences in the social world. A basic assumption is that individuals are actively engaged in interpreting events, objects, and people in their lives. The phenomena explored by IPA are usually of some personal significance to the participants (e.g., life events, relationships, or phenomena encountered in life). IPA researchers attempt to understand what it is like to stand in the shoes of the subject.~\cite{smith2011evaluating, smith2015qualitative, pietkiewicz2014practical, brocki2006critical}

The foundation of IPA was first described and conceptualized in the mid-90s~\cite{smith1997interpretative, jonathan2009interpretative}. Early in the development, IPA was mainly used in health psychology. However, IPA has rapidly grown and become one of the best known and most commonly used qualitative methodologies in psychology~\cite{smith2011evaluating, willig2007reflections}. Recently, it has branched out into other applied psychologies, e.g., clinical, counseling, educational, and occupational~\cite{jonathan2009interpretative}.

IPA draws upon and has its theoretical roots in the fundamental principles of phenomenology, hermeneutics, and ideography~\cite{pietkiewicz2012praktyczny}.

Rather than focusing on describing phenomena according to scientific criteria, phenomenological studies focus on how people perceive and talk about objects and events. It is primarily concerned with attending to the way things appear to individuals as experiences~\cite{smith2011evaluating, pietkiewicz2012praktyczny}. 

A key concept in phenomenology is lifeworld~\cite{makkreel1982husserl}. The concept, which holds dual components as it is both personal and intersubjective, is indeed multifaceted and complex, and we thus cannot claim to give it full justice in this paper. The lifeworld comprises the world of objects around us as we perceive them and our experience of our self, our body, and our relationships. It is the world which people can experience together, i.e., the common ground that we can share~\cite{moran2012husserl}. An individual's lifeworld consists of the beliefs that form hers or his everyday attitude towards herself or himself, the objective world and other people.

According to phenomenology, to get a clear view of the lifeworld and gain a true understanding of a phenomenon, it is necessary to disregard (bracket) preconceptions and judgments. This process is known as \textit{epoché} or phenomenological reduction. Through this, phenomenology researchers can better uncover what the essential and unique components are that form a given phenomenon.~\cite{pietkiewicz2012praktyczny}

While phenomenology uncovers meanings, hermeneutics interprets that meaning~\cite{backstrom2007meaning}. IPA research requires a two-stage interpretation process in which a subject is trying to make sense of their world, and the researcher, in turn, is trying to make sense of the subject trying to make sense of their world. This process has been called double hermeneutic~\cite{smith2011evaluating}. It requires an engagement and interpretation for the researcher, which connects IPA to a hermeneutic perspective.

The ideographic component of IPA refers to an in-depth analysis of single cases and individual perspectives. IPA focuses on the particular, rather than the general or universal~\cite{pietkiewicz2012praktyczny}. It involves the detailed approach to each case, followed by the search for patterns across the cases.~\cite{smith2011evaluating}. The aim of an IPA study is to present the perceptions and understandings of a particular phenomenon in detail, rather than prematurely make more general claims.

IPA studies usually utilize small, purposely selected, sample sizes, partially because the transcripts require a lot of analysis efforts. IPA researchers strive for a fairly homogeneous sample extracted from a closely defined group for whom the research question is significant and whom, usually, have an understanding of the topic~\cite{larkin2012interpretative, pietkiewicz2012praktyczny}.

The research questions in IPA studies are usually framed broadly and openly, since the intent is exploratory rather than explanatory~\cite{larkin2012interpretative}. IPA is a suitable method when one is trying to find out how individuals perceive a particular situation they are facing, and how they are making sense of their personal and social world~\cite{smith2015qualitative}. It is especially useful when one is concerned with complexity, process, or novelty~\cite{brocki2006critical}.

IPA is particularly well-suited for understanding the cognitive, emotional, and perceptual aspects of individual experiences. In relation to software engineering, we suggest that IPA would be the preferred choice when exploring the software engineers' individual account on several tasks. In IPA allows for in-depth exploration of the subjective experiences of software engineers, project managers, and other stakeholders involved in the development process. Gaining insights into their perspectives and thought processes enables researchers to better understand the challenges faced and identify potential solutions that take into account the human aspects of software development.

Today, as a result of agile transformation, software engineering organizations often emphasize and focus on groups. In fact, the group has replaced the individual as the most important entity. IPA could thus be used to counterbalance the focus and provide a detailed and nuanced description of an individual experience. Examples of possible questions that could be addressed using IPA are "How does a software engineer experience the transition from university to working life?" and "What does organizational loyalty mean to a software developer?".

We provide three examples of how research in software engineering could use IPA:

Software development team dynamics: IPA could be employed to investigate the experiences of team members working in agile environments. Exploring the intricacies of interpersonal relationships, communication preferences, and the impact of team dynamics on individuals' experiences can provide valuable insights for improving collaboration, communication, and overall performance.

Adoption of new technologies: IPA could be utilized to examine the experiences of software engineers who are learning and adopting new technologies, frameworks, or programming languages. By delving into the emotions, motivations, and cognitive processes associated with technology adoption, researchers can better understand the barriers and facilitators to learning and the overall impact on performance and job satisfaction.

Developer mental health: IPA could be used to investigate the experiences of software developers dealing with stress, burnout, or other mental health issues. By understanding the lived experiences of developers in distress, researchers can identify factors contributing to these problems and develop strategies to mitigate them, such as improved work-life balance, organizational support, and mental health resources.

We have recently started employing IPA in the software engineering domain. Only one of these studies is yet publicly available, namely Weise's MSc thesis entitled ``Frictions in software development : an interpretive phenomenological analysis''~\cite{weise2021Frictions}. In his thesis, supervised by one of the authors of the present paper, Weise explores the frictions---defined as issues that decelerate the progress of workers but do not lead to complete stagnancy---in the software development sector within a team of data analysts in the consumer electronics sector. Driven by a prior version of our guidelines to form his research design, Weise interviewed five participants working at the same company to identify causes of frictions and how they experienced them. The thesis reports on the whole IPA elements, including reflexivity.

A recent paper by Singh \& Strobel~\cite{singh2023Exploring}, published while this paper was in submission, reports on a phenomenological study that explores the lived experiences of agile developers with daily standup meetings. The study aims to provide a rich description of the essence of individual developers' experiences (19 participants) with daily standup meetings and to better understand the phenomenon overall. The authors employed IPA to present the participants' conversations as carefully extracted phenomenological examples to better understand the phenomenon of standup meetings. The authors found that developers had varying experiences with daily standup meetings, ranging from positive to negative. Some developers found daily standup meetings helpful for communication and collaboration, while others found them time-consuming and unproductive. A finding that was perhaps brought by the chosen methodology was that junior developers were afraid of standup meetings, but on the other hand, they experienced information sharing as learned opportunities. The authors provide several implications for practice, but, in summary, they suggest that organizations should consider the individual needs and preferences of their developers when implementing daily standup meetings.

\subsubsection{Narrative Analysis}
In an experiment conducted in the 1940s, psychologists Fritz Heider and Marianne Simmel demonstrated the importance of stories to humans~\cite{heider1944experimental}. In the experiment, the participants were shown a sequence of pictures that included abstract shapes such as squares, triangles, and lines. When the participants later were asked to describe the pictures, they replied by telling short stories.

According to Murray~\cite{murray2003narrative}, humans live their lives through stories and describe their experiences and their selves in terms of stories. Humans need stories to make sense of their lived experience, as they can help to connect their past, present, and future. It allows them to maintain a coherent self-identity~\cite{kugelmann2001introducing, ricoeur2010time}. It is through the use of stories that we define who we are, were and how we will be in the future.

Narrative research~\cite{riessman2005narrative, kohler2000analysis} is the study of stories. It seeks to uncover how humans make sense of an ever-changing world, based on a belief that it is through a narrative that we can bring a sense of order to the seeming disorder in our world~\cite{murray2003narrative}. It is often focused on life experiences of a single event or a series of events for a few individuals~\cite{creswell2013research}.

A pioneer of narrative research is the American psychologist Theodore R. Sarbin. The term narrative psychology was introduced in 1986 in his book Narrative Psychology: The storied nature of human conduct~\cite{sarbin1986narrative}. Sarbin claimed that human behaviors are best explained using stories and that narrative should be a root metaphor in psychology. He also argued that narratives should be identified through qualitative research~\cite{crossley2000introducing}.

Narrative research examines how people construct their self-accounts, and it is often used for scrutinizing how people manage their different senses of self~\cite{burck2005comparing}. It can, however, also lend itself to a global view of human experiences.

Moreover, narrative research data can be anything that provides information and details to a contextualized story. Examples of data are observations, diaries, letters, interviews, artifacts, and photographs~\cite{petty2012ready}. Still, the primary data source is the interview~\cite{murray2003narrative}.

One type of narrative interview is the so-called life-story interview, which aims to capture an extended account of the participants' lives. These types of interviews are complicated, and several interview occasions are often needed in order for the participant to feel secure enough to reflect on her/his life experiences.~\cite{murray2003narrative} Another interview type is the episodic interview~\cite{flick2014introduction} where the participants are encouraged to tell stories about particular experiences or disruptive episodes in their lives.

According to Murray~\cite{murray2003narrative}, the analysis of narrative accounts can be divided into two broad phases, i.e., the descriptive phase and the interpretive phase. In the first phase, the researchers briefly summarize the narratives, identify their beginning, middle, and end, and capture their overall meaning and any particular issues raised by them. In the second phase, the researchers go beyond the descriptive and connect the narrative with the broader theoretical literature that is being used to interpret the story. This phase thus requires a simultaneous and deep intimacy with the narrative accounts and with the relevant literature.

In narrative studies, the collected data and the analyzed result may not only answer a research question. Even if a narrative inquiry is focused on a particular experience, it often reveals additional aspects of life that are not identified as the primary focus of the study~\cite{overcash2003narrative}. Therefore, the research questions shall be detailed enough to provide guidance in the research, but they should, at the same time, include a high degree of flexibility. Often, the research questions of narrative inquiry are refined during the research process as more insights of the phenomena are gained~\cite{shkedi2005multiple, connelly2000narrative}.

Narrative analysis focuses on the stories people tell about their experiences, emphasizing the construction, organization, and meaning of these narratives. In software engineering research, narrative analysis can reveal the underlying values, beliefs, and motivations of individuals involved in the development process. By exploring how these narratives are constructed and how they reflect the sociocultural context, researchers can gain a more in-depth understanding of the dynamics shaping software engineering practices and identify ways to enhance them.

As an example, software development has, in the past twenty years, gone from being an individual occupation to becoming an occupation that requires teamwork and that emphasizes collaboration and cooperation. We think that it would be interesting to use a narrative inquiry to get insights into a software developer's experiences of this major transformation process. In addition, we believe that the narrative method would be helpful in understanding the software engineer's professional identity~\cite{koller2012analyse}.%

Project success and failure stories: Researchers can use narrative analysis to examine stories of successful and failed software development projects, shedding light on the factors that contribute to these outcomes. By exploring the narrative structure and themes, researchers can identify patterns in the decision-making processes, organizational culture, and team dynamics that may influence project success or failure.

Developer career trajectories: Analyzing the narratives of software developers' career paths can provide insights into factors influencing their professional growth, job satisfaction, and retention. This information can help organizations develop strategies to attract, retain, and support their software engineering talent. By examining the recurring themes, turning points, and career-defining moments in developers' stories, researchers can gain a more profound understanding of the factors that drive career choices and satisfaction.

Organizational culture: Narrative analysis can be employed to explore stories about the culture of software development organizations. These insights can reveal the values and norms that impact the work environment and the overall satisfaction and performance of the employees. By identifying the key elements that shape organizational narratives, researchers can inform interventions to promote a positive and supportive work culture.

\subsubsection{Discourse Analysis}
Discourse analysis is a broad term for the many traditions and methods by which discourse may be identified, defined, and analyzed~\cite{morgan2010discourse}. Discourse analysis is related to grammar analysis, but there are differences. Grammar analysis includes observation of sentence structure, word usage, and stylistic choices on the sentence level. Even if such analysis might include entities like culture, its focus is not human spoken discourse. Discourse analysis, in turn, observes the conversational, cultural, and use of language by its native population. %

In discourse analysis, the words themselves are virtually meaningless to us. It is through the shared, mutually agreed use of language that meaning is created. The language shapes our understanding of reality and defines the creation and maintenance of social norms, the construction of personal and group identities, and the negotiation of social and political interaction~\cite{crowe1998power, gee2014introduction, lyons1968introduction, chandler2007semiotics, starks2007choose}.

In a broad sense, discourses are defined as systems of meaning that are related to the interactional and wider sociocultural context and operate regardless of the speakers’ intentions~\cite{georgaca2012discourse}. It should, however, be noted that there are many traditions within discourse analysis and that these use their own, slightly different, definitions. There also exists different views as to what degree the individual is an actor in forming discourse or being influenced by existing discourses. As an example, within the Foucauldian research, discourse is defined as a group of statements, objects, or events that represent knowledge about, or construct, a particular topic. Here, language is viewed as a social performance or a social action. It both creates social phenomena and is representative of social phenomena. %

Given its emphasis on construction and function, discourse analysis does not make claims about the reality of people’s lives or experiences. Instead, it examines the ways in which reality and experience are constructed through social and interpersonal processes through language~\cite{georgaca2012discourse, starks2007choose}. Discourse studies often do not provide a precise answer to a specific problem. Instead, they provide an understanding and a clarification of the essence of the problem and the underlying assumptions that enable its existence. They thus present a deeper and exhaustive view of the problem and how we are affected. In addition, discourse analysis can be used to reveal implicit and unacknowledged aspects of human behavior, and, for example, making salient either latent or dominant discourses in society. 

Discourse analysis can be applied to any type of text, i.e., to anything that has meaning. However, most studies tend to analyze written or spoken language~\cite{georgaca2012discourse, parker2015critical}. Regarding data sampling and size, discourse analyses often rely on relatively small numbers of participants or texts. Partially because that analysis is very labor-intensive and large amounts of data would be prohibitive~\cite{georgaca2012discourse}; however, the sample size ultimately depends on the study objective. It is possible to analyze in-depth a single participant and compare it with written documents. On the other hand, if the objective is to understand variations in used language across individuals or settings, a larger size is required.~\cite{starks2007choose} The discourse is independent of context and exists even if it is defined in a conversation in the sauna or in the boardroom.

Discourse analysis examines the ways in which language is used to construct meaning and shape social interactions. In software engineering research, discourse analysis can be valuable in understanding communication patterns, power dynamics, and the construction of knowledge among stakeholders. By examining the linguistic and rhetorical devices used by individuals involved in software development, researchers can gain insights into the social and cultural contexts that shape software engineering practices.

We recognize that discourse analysis could add value and insights to the research area of software engineering. For example, we believe that it would be a viable option to generate knowledge regarding the transformation of the meaning of the agile concept. Even though the agile approach has a definition and has been around for well over a decade, we argue that the values and beliefs that govern the concept have changed over time and this varies between different organizational roles. A more profound understanding of agile concepts would, for example, provide valuable information that could be used to improve large organizations' transition to an agile methodology. 

In addition, we think that discourse analysis could be used to understand the relation of power in agile software organizations. It could possibly shed light on the relationship between discourse and power, e.g., what discourse exists in software engineering organizations, and how do they empower some groups while dis-empowering others?

We provide three examples of using discourse analysis in software engineering:

Communication in remote teams: Discourse analysis can be used to study the communication patterns among remote software development teams. By examining the use of language, communication channels, and interaction styles, researchers can identify potential issues, such as misunderstandings or miscommunications, and develop best practices for effective communication in a distributed work environment.

Decision-making processes: By analyzing the discourse during meetings and discussions, researchers can explore the decision-making processes in software development teams. This understanding can aid in developing strategies to enhance collaborative decision-making, reduce potential conflicts, and promote a more inclusive decision-making environment that considers diverse perspectives and opinions.

Gender and diversity in software engineering: Discourse analysis can be employed to examine the language used in software engineering contexts, revealing implicit biases and power dynamics related to gender, race, and other aspects of diversity. By scrutinizing the ways in which language is used to construct and maintain social hierarchies, researchers can inform interventions to promote a more inclusive and diverse software engineering community.

\highlight{The three qualitative methods that we report in the present paper are mostly of interest to the natural settings of the ABC framework. While they can potentially also add value to qualitative judgment studies and sample studies of the neutral setting, the methods emphasize the maximization of the C in the ABC, that is realism of context, rather than generalizing over actors or behaviors. However, they also allow for increased precision in understanding the behavior of actors and in that sense can add value also in other settings, whether neutral or contrived.}

\subsubsection{Limitations and challenges}
Many limitations and challenges are shared among the three described methods, and also with qualitative methods in general. Firstly, there are ethical challenges that the researchers must consider. The central principles of research ethics are informed consent, confidentiality, and avoidance of harm, which all create dilemmas for qualitative studies~\cite{houghton2010ethical}.

Qualitative inquiries are discovery-oriented, and it is hard to anticipate upfront what discoveries of human behavior, thoughts, or feelings will emerge. In such situations, informed consent becomes complicated since researchers cannot predict the scope of the study. To formulate a description that offers participants a comprehensive account of their experience of the study clearly becomes challenging~\cite{mcleod1996qualitative, houghton2010ethical}. This is certainly true for narrative analysis and also for IPA, where the boundaries are loose, and the researchers are encouraged to have broad and inclusive research questions.

In addition, and in particular for narrative analysis, confidentiality can form a major challenge. In a narrative inquiry, the stories that unfold during interviews are likely to be unique, holding several clues and markers that could possibly identify the interviewee, which makes confidentiality challenging to fulfill. The risk could to some extent be mitigated by allowing informants to read pre-publication drafts before publication. There are, however, limits to this procedure, since the informants may not fully appreciate what may happen when the publication enters the public domain~\cite{mcleod1996qualitative}.

The third ethical consideration, i.e., avoidance of harm, involves predicting the risk-benefit ratio of the research, which, in qualitative research, often is difficult~\cite{cutcliffe2002leveling}. Yet again, since narrative inquiry and IPA are both flexible methods, it is hard to control the direction of the interview, which makes it hard for the researcher to foresee the result of  interview questions and avoid potentially stressful situations for the interviewee.

Second, the nature of data collection in the three methods, observation, and interviews alike, raises ethical issues related to how the relationships are formed and managed, to the nature of the power, and to how the relationship affects the participants~\cite{ginger2004toward, houghton2010ethical}. There is also a high risk that the narrative and IPA style of the interview becomes transformative or even therapeutic. This leaves a heavy burden of responsibility on the shoulders of the researcher, especially when the topic is sensitive to the interviewee~\cite{liamputtong2013qualitative, stuhlmiller2001narrative, hunter2010analysing}.

Third, a common critique of IPA and narrative analysis is their dependability of language. The informants may not always be able to accurately convey the subtleties and nuances of their experience~\cite{willig2013introducing}. According to Smith et al.~\cite{smith1997interpretative}, this could be managed by a professional researcher who can interpret the participant's emotional state and ask follow-up questions. 
The language dependability is also a challenge for the researcher. It can be difficult to put into words the rich knowledge extracted from qualitative inquiry, which may hold information that is subtle, hidden, and contextually bound~\cite{kapoulas2012understanding}.
In addition, there are also inherent ambiguities in human language that need to be recognized by the researcher in the analysis~\cite{atieno2009analysis}. As an example, the word blue could signify the color, a political orientation or a state of mind.

Fourth, another challenge shared among the three methods is their process flexibility and their researcher dependence. Out of all the elements and processes of qualitative research, the analysis one is often the most sensitive~\cite{kapoulas2012understanding}. IPA and narrative analysis do not provide any formula for how to decide what parts of the data should be highlighted. What to emphasize thus relies heavily and solely on the researcher and her or his experience and knowledge~\cite{wiles2005narrative}. This implies that different conclusions can be derived based on the same information, depending on the personal characteristics of the researcher\cite{maxwell2012qualitative}. 

Moreover, even if IPA includes a comprehensively described process consisting of several well-defined steps, the researchers are encouraged to engage it with flexibility and adapt the process to the phenomena under investigation~\cite{smith1997interpretative}. This flexibility and the engaged role of the researcher bring potential challenges to IPA studies~\cite{cronin2015brief}. IPA has been criticized for not providing guidelines on how to incorporate reflexivity (see section~\ref{section_reflexivity}) into the process and for not specifying how researcher conceptions influence analysis. As a researcher, it might be difficult to keep the balance between being fully engaged while, at the same time, remaining unbiased~\cite{foster2010adolescents, clancy2013reflexivity}.

Method related flexibility is also challenging in discourse analysis. Generally, proponents of discourse analysis believe that meaning is never fixed and everything is therefore always open to interpretation and negotiation. Moreover, the vast number of options available through the various traditions might cause method problems, since each tradition has its own epistemological position, concepts, procedures, communication, and a particular understanding of discourse~\cite{morgan2010discourse}.

Moreover, the aim of a narrative inquiry is not to find one generalized truth, but rather many truths or narratives. Since these narratives are created between the participant and the researcher in a particular social and cultural context, it raises issues about whether research findings can be seen as valid. However, it has also been argued that if a phenomenon exists in one setting, is it plausible to believe that it also exists in others~\cite{polkinghorne2007validity, hunter2010analysing}. 

Finally, an overall limitation of our contribution lies in our reasoning and vision that the three introduced methods, reflexivity practices, and quality criteria are fully applicable, and fruitfully in the software engineering context. We could argue for the positive sides of these concepts, we provided examples of potential applications in our field, and we highlighted some examples of published work in the area. Yet, we recognize that there is not enough evidence in the form of published studies that this is the case. On the one hand, this was the main motivating factor that drove us to develop these guidelines. On the other hand, we call our peers to engage with the new material brought by the present paper and report on its applicability, including the extent to which it is applicable in software engineering research. The ABC framework of software engineering has guided us to frame the applicability of COREQ in a wide range of software engineering studies in the neutral and the natural settings. Furthermore, reflexivity as a practice, and the three research methods we are suggesting to software engineering research, allow for maximizing realism for context in studies, and mostly apply to natural settings studies.

\section{Conclusion}
\label{lbl_conclusion}

The purpose of this study was to extend the community's current body of knowledge regarding available qualitative methods and their quality assurance frameworks, and to provide recommendations and guidelines for their use. Our overarching research question was \textit{\resq}.

Supported by the results of a literature review and an epistemological argument, we conclude that future qualitative studies would benefit from:

\begin{enumerate}
    \item adopting a broader set of qualitative research methods,
    \item emphasizing reflexivity, 
    \item and employing qualitative guidelines and quality criteria.
\end{enumerate}

Three qualitative methods frequently used by social science researchers to explore organizational life, and that potentially also could add value to software engineering research, are interpretative phenomenological analysis, narrative analysis, and discourse analysis.

Moreover, we argue that reflexivity is highly important in software engineering studies since the researchers often have preconceived opinions of the phenomena under investigation and that the risk of bias therefore is high.

Finally, we recommend qualitative software engineering researchers to utilize quality criteria, such as the COREQ checklist, but we do not discourage the use of the Big-Tent criteria as guidance when conducting their studies either, and to reflect on their epistemological stance.

\section{Data Availability Statement}

The data that supports the findings of this study are available in the supplementary material of this article (Appendix).

\section{Acknowledgments}
We acknowledge the support of Swedish Armed Forces, Swedish Defense Materiel Administration and Swedish Governmental Agency for Innovation Systems (VINNOVA) in the project number 2017-04874. Daniel Graziotin was supported by the Alexander von Humboldt (AvH) Foundation. We are grateful to Katharina Plett for proofreading the manuscript.

\bibliography{references}

\section{Appendix A}
\begingroup
\footnotesize
\begin{longtabu}{ | p{0.90\textwidth} | } \hline
\textbf{Included Papers} \\
\hline
The Influence of Technical Debt on Software Developer Morale	\cite{besker2020influence}	\\	\hline
Code and Commit Metrics of Developer Productivity: a Study on Team Leaders Perceptions	\cite{oliveira2020code}	\\	\hline
Quality Attribute Trade-offs in the Embedded Systems Industry: an Exploratory Case Study Open Access	\cite{sas2019quality}	\\	\hline
Regression Testing for Large-scale Embedded Software Development – Exploring the State of Practice	\cite{minhas2020regression}	\\	\hline
The Third Design Space: a Postcolonial Perspective on Corporate Engagement With Open Source Software Communities Open Access	\cite{kendall2020third}	\\	\hline
A Generational Perspective on the Software Workforce: Precocious Users of Social Networking in Software Development	\cite{ghobadi2020generational}	\\	\hline
Women in Coding Boot Camps: an Alternative Pathway to Computing Jobs	\cite{lyon2020women}	\\	\hline
Requirements Specification for Developers in Agile Projects: Evaluation by Two Industrial Case Studies	\cite{medeiros2020requirements}	\\	\hline
An Agile-based Integrated Framework for Mobile Application Development Considering Ilities	\cite{martinez2020agile}	\\	\hline
Data-driven and Tool-supported Elicitation of Quality Requirements in Agile Companies	\cite{oriol2020data}	\\	\hline
From Female Computers to Male Computers: or Why There Are So Few Women Writing Algorithms and Developing Software	\cite{tassabehji2020female}	\\	\hline
Designing Smart City Mobile Applications: an Initial Grounded Theory	\cite{farias2019designing}	\\	\hline
Adopting Devops in the Real World: a Theory, a Model, and a Case Study	\cite{luz2019adopting}	\\	\hline
A Modeling Approach for Systems-of-systems by Adapting Iso/iec/ieee 42010 Standard Evaluated by Goal-question-metric	\cite{chaabane2019modeling}	\\	\hline
Integrating Ux Principles and Practices Into Software Development Organizations: a Case Study of Influencing Events	\cite{kashfi2019integrating}	\\	\hline
Breaking the Flow: a Study of Contradictions in Information Systems Development (isd)	\cite{dennehy2019breaking}	\\	\hline
An Exploratory Study for Investigating the Issues and Current Practices of Service-oriented Architecture Adoption	\cite{hamzah2019exploratory}	\\	\hline
Categorizing the Content of Github Readme Files	\cite{prana2019categorizing}	\\	\hline
Example-driven Modeling: on Effects of Using Examples on Structural Model Comprehension, What Makes Them Useful, and How to Create Them	\cite{zayan2019example}	\\	\hline
A Model of Requirements Engineering in Software Startups	\cite{melegati2019model}	\\	\hline
The Current State of Software License Renewals in the I.t. Industry	\cite{ghosh2019current}	\\	\hline
Misaligned Values in Software Engineering Organizations	\cite{lenberg2019misaligned}	\\	\hline
Alone or Together? Inter-organizational Affiliations of Open Source Communities	\cite{eckert2019alone}	\\	\hline
Towards Organisational Learning Enhancement: Assessing Software Engineering Practice	\cite{fannoun2019towards}	\\	\hline
Agile Self-selecting Teams Foster Expertise Coordination	\cite{rejab2019agile}	\\	\hline
Information Flow in Software Testing - an Interview Study With Embedded Software Engineering Practitioners	\cite{strandberg2019information}	\\	\hline
Factors and Actors Leading to the Adoption of a Javascript Framework	\cite{pano2018factors}	\\	\hline
System Requirements-oss Components: Matching and Mismatch Resolution Practices – an Empirical Study	\cite{ayala2018system}	\\	\hline
Work Practices and Challenges in Continuous Integration: a Survey With Travis Ci Users	\cite{pinto2018work}	\\	\hline
Coordination Challenges in Large-scale Software Development: a Case Study of Planning Misalignment in Hybrid Settings	\cite{bick2017coordination}	\\	\hline
On User Rationale in Software Engineering	\cite{kurtanovic2018user}	\\	\hline
Quality of Software Requirements Specification in Agile Projects: a Cross-case Analysis of Six Companies	\cite{medeiros2018quality}	\\	\hline
The Influencing Causes of Software Unavailability: a Case Study from Industry	\cite{ebad2018influencing}	\\	\hline
Transition of Organizational Roles in Agile Transformation Process: a Grounded Theory Approach	\cite{jovanovic2017transition}	\\	\hline
User Satisfaction and System Success: an Empirical Exploration of User Involvement in Software Development	\cite{bano2017user}	\\	\hline
Refining a Model for Sustained Usage of Agile Methodologies	\cite{senapathi2017refining}	\\	\hline
Quality Attribute Trade-offs in the Embedded Systems Industry: an Exploratory Case Study Open Access	\cite{sas2019quality}	\\	\hline
Team-external Coordination in Large-scale Software Development Projects	\cite{sablis2020team}	\\	\hline
Factors Influencing the Adoption of Iso/iec 29110 in Thai Government Projects: a Case Study	\cite{siddoo2017factors}	\\	\hline
Cheap Talk, Cooperation, and Trust in Global Software Engineering: an Evolutionary Game Theory Model With Empirical Support	\cite{wang2016cheap}	\\	\hline
Improving the Delivery Cycle: a Multiple-case Study of the Toolchains in Finnish Software Intensive Enterprises	\cite{makinen2016improving}	\\	\hline
An Empirical Study and a Framework for Effective Risk Management in Scrum	\cite{aliempirical}	\\	\hline
Artefacts and Agile Method Tailoring in Large-scale Offshore Software Development Programmes	\cite{bass2016artefacts}	\\	\hline
Multi-level Agile Project Management Challenges: a Self-organizing Team Perspective	\cite{hoda2016multi}	\\	\hline
Software Development in Startup Companies: the Greenfield Startup Model	\cite{giardino2015software}	\\	\hline
The Daily Stand-up Meeting: a Grounded Theory Study	\cite{stray2016daily}	\\	\hline
Professional Identity Construction Among Software Engineering Students: a Study in India	\cite{mishra2016professional}	\\	\hline
Toward an Agile Approach to Managing the Effect of Requirements on Software Architecture During Global Software Development	\cite{alsahli2016toward}	\\	\hline
A Qualitative Study on Debugging Under an Enterprise Ide	\cite{zayour2016qualitative}	\\	\hline
A Multiple Case Study on the Inter-group Interaction Speed in Large, Embedded Software Companies Employing Agile	\cite{martini2016multiple}	\\	\hline
Customer Interaction in Software Development: a Comparison of Software Methodologies Deployed in Namibian Software Firms	\cite{iyawa2016customer}	\\	\hline
Adopting Agile Software Development: the Project Manager Experience	\cite{taylor2016adopting}	\\	\hline
Agile Transition and Adoption Human-related Challenges and Issues: a Grounded Theory Approach	\cite{gandomani2016agile}	\\	\hline
Lessons Learned from Applying Social Network Analysis on an Industrial Free/libre/open Source Software Ecosystem	\cite{teixeira2015lessons}	\\	\hline
A Conceptual Framework of Challenges and Solutions for Managing Global Software Maintenance	\cite{ulziit2015conceptual}	\\	\hline
An Empirically-developed Framework for Agile Transition and Adoption: a Grounded Theory Approach	\cite{gandomani2015empirically}	\\	\hline
Progressive Outcomes: a Framework for Maturing in Agile Software Development	\cite{fontana2015progressive}	\\	\hline
A Conceptual Framework to Study the Role of Communication Through Social Software for Coordination in Globally-distributed Software Teams	\cite{giuffrida2015conceptual}	\\	\hline
The Design Space of Bug Fixes and How Developers Navigate It	\cite{murphy2014design}	\\	\hline
Perceived Occupational Stressors and the Health So Ware Professionals in Bengaluru, India	\cite{anantharaman2018}	\\	\hline
Social Debt in Software Engineering: Insights from Industry	\cite{tamburri2015social}	\\	\hline
\hline
\caption{Papers included in the literature review.}
\label{table:slr_studies}
\end{longtabu}
\endgroup

\section{Appendix B}
\begingroup
\footnotesize
\begin{longtabu}{ | p{0.40\textwidth} | p{0.40\textwidth} | } \hline
\\
\textbf{Domain 1: Research team and reflexivity} 23\% & \textbf{Domain 2: Study design} 38\% \\
\emph{A: Personal Characteristics} 29\% & \emph{C: Theoretical framework} 40\% \\ 
1. Interviewer/facilitator 21\% & 9. Methodological orientation and theory 70\% \\ 
2. Credentials 79\% & \\ 
3. Occupation 16\% & \emph{D: Participant selection} 50\% \\ 
4. Gender 18\% & 10. Sampling 63\% \\
5. Experience and training 10\% & 11. Method of approach 40\% \\
 & 12. Sample size 95\% \\
\emph{B: Relationship with participants} 14\% & 13. Non-participation 2\% \\
6. Relationship established 13\% & \\
7. Participant knowledge of the interviewer 21\% & \emph{E: Setting} 44\% \\ 
8. Interviewer characteristics 7\% & 14. Setting of data collection 58\% \\
 & 15. Presence of non-participants 5\% \\ 
\textbf{Domain 3: Analysis and findings} 30\% & 16. Description of sample 68\% \\ 
\emph{G: Data analysis} 30\% & \\
24. Number of data coders 22\%& \emph{F: Data collection} 36\% \\
25. Description of the coding tree 20\% & 17. Interview guide 60\% \\ 
26. Derivation of themes 72\% & 18. Repeat interviews 9\% \\ 
27. Software 30\% & 19. Audio/visual recording 66\% \\
28. Participant checking 4\% & 20. Field notes 12\% \\
 & 21. Duration 62\% \\
 & 22. Data saturation 34\% \\
 & 23. Transcripts returned 7\% \\
\hline
\caption{Empirical overview of the result for the quality indicator properties defined by COREQ~\cite{tong2007consolidated}.}
\label{table:result_RRB}
\end{longtabu}
\endgroup

\section{Appendix C}

\begingroup
\footnotesize
\begin{longtabu}{ |p{0.40\textwidth} | p{0.40\textwidth} | } 
\hline
\textbf{Criterion} & \textbf{Guiding question(s)} \\
\hline
\textbf{Domain 1: Research team and reflexivity} & \\
\emph{A: Personal Characteristics} & \\ 
1. Interviewer/facilitator & Which author/s conducted the interview or focus group? \\ 
2. Credentials & What were the researcher's credentials? \\ 
3. Occupation & What was their occupation at the time of the study? \\ 
4. Gender & Was the researcher male or female? \\
5. Experience and training & What experience or training did the researcher have? \\
 & \\
\emph{B: Relationship with participants} & \\
6. Relationship established & Was a relationship established prior to study commencement? \\ 
7. Participant knowledge of the interviewer & What did the participants know about the researcher? \\
8. Interviewer characteristics & What characteristics were reported about the interviewer/facilitator? Bias? \\ 
\hline
\textbf{Domain 2: Study design} & \\ 
\emph{C: Theoretical framework} & \\ 
9. Methodological orientation and theory & What methodological orientation was stated to underpin the study? \\
 & \\
\emph{D: Participant selection} & \\ 
10. Sampling & How were participants selected? \\
11. Method of approach & How were participants approached? \\
12. Sample size & How many participants were in the study? \\ 
13. Non-participation & How many people refused to participate or dropped out? \\
 & \\
\emph{E: Setting} & \\ 
14. Setting of data collection & Where was the data collected? \\ 
15. Presence of non-participants & Was anyone else present besides the participants and researchers? \\
16. Description of sample & What are the important characteristics of the sample? \\
 & \\
\emph{F: Data collection} & \\ 
17. Interview guide & Were questions, prompts, guides provided by the authors? Was it pilot tested? \\
18. Repeat interviews & Were repeat interviews carried out? \\
19. Audio/visual recording & Did the research use audio or visual recording to collect the data? \\ 
20. Field notes & Were field notes made during and/or after the interview or focus group? \\ 
21. Duration & What was the duration of the interviews or focus group? \\
22. Data saturation & Was data saturation discussed? \\
23. Transcripts returned & Were transcripts returned to participants for comment and/or correction? \\
\hline
\textbf{Domain 3: Analysis and findings} & \\ 
\emph{G: Data analysis} & \\ 
24. Number of data coders & How many data coders coded the data? \\
25. Description of the coding tree & Did authors provide a description of the coding tree? \\
26. Derivation of themes & Were themes identified in advance or derived from the data? \\
27. Software & What software, if applicable, was used to manage the data? \\
28. Participant checking & Did participants provide feedback on the findings? \\
 & \\
\emph{H: Reporting} & \\
29. Quotations presented & Were participant quotations presented to illustrate the themes/findings? \\
30. Data and findings consistent & Was there consistency between the data presented and the findings? \\
31. Clarity of major themes & Were major themes clearly presented in the findings? \\
32. Clarity of minor themes & Is there a description of diverse cases or discussion of minor themes? \\
\hline
\caption{COREQ Quality criteria~\cite{tong2007consolidated}}
\label{table:coreq}
\end{longtabu}
\endgroup 

\section{Appendix D}

\begingroup
\footnotesize

\begin{longtable}{|l|l|}
\caption{Papers that fulfilled each of the 28 first COREQ criteria\label{table:coreq2papers}}\\ 
\hline
\textbf{Criterion}                                                                                                                         & \textbf{Count}  \endfirsthead 
\hline
\textbf{A. Characteristics}                                                                                                                &                 \\ 
\hline
Which author/s conducted the interview or focus group?                                                                                     & 13              \\
What were the researcher’s credentials? e.g., PhD, MD                                                                                      & 48              \\
What was their occupation at the time of the study?                                                                                        & 10              \\
Was the researcher male or female?                                                                                                         & 11              \\
What experience or training did the researcher have?                                                                                       & 6               \\ 
\hline
\textbf{B. Relationship}                                                                                                                   &                 \\ 
\hline
Was a relationship established prior to study commencement?                                                                                & 8               \\
What did the participants know about the researcher? e.g., personal goals, reasons for doing the research                                  & 13              \\
What characteristics were reported about the interviewer/facilitator? e.g., Bias, assumptions, reasons and interests in the research topic & 4               \\ 
\hline
\textbf{C. Framework}                                                                                                                      &                 \\ 
\hline
What methodological orientation was stated to underpin the study?                                                                          & 43              \\ 
\hline
\textbf{D. Selection}                                                                                                                      &                 \\ 
\hline
How were participants selected? e.g., purposive, convenience, consecutive, snowball                                                        & 38              \\
How were participants approached? e.g., face-to-face, telephone, mail, email                                                               & 24              \\
How many participants were in the study?                                                                                                   & 57              \\
How many people refused to participate or dropped out? Reasons?                                                                            & 1               \\ 
\hline
\textbf{E. Setting}                                                                                                                        &                 \\ 
\hline
Where was the data collected? e.g., home, clinic, workplace                                                                                & 35              \\
Was anyone else present besides the participants and researchers?                                                                          & 3               \\
What are the important characteristics of the sample? e.g., demographic data, date                                                         & 41              \\ 
\hline
\textbf{F. Collection}                                                                                                                     &                 \\ 
\hline
Were questions, prompts, guides provided by the authors? Was it pilot tested?                                                              & 35              \\
Were repeat interviews carried out? If yes, how many?                                                                                      & 5               \\
Did the research use audio or visual recording to collect the data?                                                                        & 38              \\
Were field notes made during and/or after the interview or focus group?                                                                    & 7               \\
What was the duration of the interviews or focus group?                                                                                    & 36              \\
Was data saturation discussed?                                                                                                             & 20              \\
Were transcripts returned to participants for comment and/or correction?                                                                   & 4               \\ 
\hline
\textbf{G. Analysis}                                                                                                                       &                 \\ 
\hline
How many data coders coded the data?                                                                                                       & 12              \\
Did authors provide a description of the coding tree?                                                                                      & 11              \\
Were themes identified in advance or derived from the data?                                                                                & 39              \\
What software, if applicable, was used to manage the data?                                                                                 & 16              \\
Did participants provide feedback on the findings?                                                                                         & 2               \\
\hline
\end{longtable}

\endgroup

\end{document}